\documentstyle[11pt]{article}
\begin{document} 
\thispagestyle{empty} 
\baselineskip 20pt 
\rightline{UOSTP-99-012} 
\rightline{SNUTP-99-050} 
\rightline{KIAS-P99106} 
\rightline{{hep-th/9912083}} 
 
\

\def\tr{{\rm tr}\,} \newcommand{\beq}{\begin{equation}} 
\newcommand{\eeq}{\end{equation}} \newcommand{\beqn}{\begin{eqnarray}} 
\newcommand{\eeqn}{\end{eqnarray}} \newcommand{\bde}{{\bf e}} 
\newcommand{\balpha}{{\mbox{\boldmath $\alpha$}}} 
\newcommand{\bsalpha}{{\mbox{\boldmath $\scriptstyle\alpha$}}} 
\newcommand{\betabf}{{\mbox{\boldmath $\beta$}}} 
\newcommand{\bgamma}{{\mbox{\boldmath $\gamma$}}} 
\newcommand{\bbeta}{{\mbox{\boldmath $\scriptstyle\beta$}}} 
\newcommand{\lambdabf}{{\mbox{\boldmath $\lambda$}}} 
\newcommand{\bphi}{{\mbox{\boldmath $\phi$}}} 
\newcommand{\bslambda}{{\mbox{\boldmath $\scriptstyle\lambda$}}} 
\newcommand{\ggg}{{\boldmath \gamma}} \newcommand{\ddd}{{\boldmath 
\delta}} \newcommand{\mmm}{{\boldmath \mu}} 
\newcommand{\nnn}{{\boldmath \nu}} 
\newcommand{\diag}{{\rm diag}} 
\newcommand{\bra}[1]{\langle {#1}|} 
\newcommand{\ket}[1]{|{#1}\rangle} 
\newcommand{\sn}{{\rm sn}} 
\newcommand{\cn}{{\rm cn}} 
\newcommand{\dn}{{\rm dn}} 
\newcommand{\bI}{{I}}  
\newcommand{\bJ}{{J}}  
\newcommand{\bmu}{{m}}  
\newcommand{\bnu}{{n}} 
\newcommand{\bp}{{p}} 
\newcommand{\bq}{{q}}

\vskip 0cm 
\centerline{\Large\bf Complete Supersymmetric Quantum Mechanics} 
\centerline{\Large\bf of Magnetic Monopoles in $N=4$ SYM Theory}

\vskip 0.2cm 
 
\vskip 1.2cm 
\centerline{\large\it 
Dongsu Bak $^a$\footnote{Electronic Mail: dsbak@mach.uos.ac.kr}, 
Kimyeong Lee $^{b}$\footnote{Electronic Mail: 
kimyeong@kias.re.kr} 
and Piljin Yi$^{b}$\footnote{Electronic Mail: 
piljin@kias.re.kr}} 
\vskip 10mm 
\centerline{ \it $^a$ Physics Department, 
University of Seoul, Seoul 130-743, Korea} 
\vskip 3mm 
\centerline{ \it $^b$ School of Physics, Korea Institute for Advanced 
Study} 
\centerline{ \it 207-43, Cheongryangryi-Dong, Dongdaemun-Gu, Seoul 
130-012, Korea 
} 
\vskip 1.2cm 
\begin{quote} 
{\baselineskip 16pt We find the most general low energy dynamics of
1/2 BPS monopoles in the $N=4$ supersymmetric Yang-Mills (SYM)
theories when all six adjoint Higgs expectation values are turned
on. When only one Higgs is turned on, the Lagrangian is purely
kinetic. When all the rest five are turned on a little bit, however,
this moduli space dynamics is augmented by five independent potential
terms, each in the form of half the squared norm of a Killing vector
field on the moduli space. A generic stationary configuration of the
monopoles can be interpreted as {\it stable non BPS dyons}, previously
found as nonplanar string webs connecting D3-branes. The
supersymmetric extension is also found explicitly, and gives the
complete quantum mechanics of monopoles in $N=4$ SYM theory.}
\end{quote}

 
\newpage 
 
\baselineskip 20pt 

\section{Introduction} 
 
Recently new stable solitons in Yang-Mills theories were constructed,  
whose electric charges and magnetic charges are not proportional to each 
other. These new solitons exist only when more than one adjoint Higgs fields  
are involved, so are natural in Yang-Mills theories with extended  
supersymmetry. Their classical aspects have been studied in the context of 
${ N}=4$ supersymmetric Yang-Mills  
theories~\cite{yi,hash,bak,kml,ioannidou,sutcliffe}. Of these dyons, 
some preserve 1/4 of ${N}=4$ supersymmetry, and thus are known as 1/4 BPS 
dyons, while others do not preserve any supersymmetry at all. 
 
It is also well known that the supersymmetric Yang-Mills theories
arise as a low energy description of parallel D3 branes in the type
IIB string theory~\cite{witten}. In this context, dyons arise as
string webs ending on D3 branes. For instance, more traditional 1/2
BPS monopoles and dyons are represented by straight $(p,q)$ string
segments ending on a pair of D3 branes, 
while a 1/4 BPS dyon corresponds to a
properly oriented, planar web of strings ending on more than two D3
branes\cite{bergman}.  The stable non-BPS states are realized as the
most general form of string web, which is typically nonplanar. Some
non-BPS dyons can be thought of as deformation of 1/4 BPS dyons which
results as D3 positions themselves get moved and become nonplanar
in the six transverse directions. (A  numerical study of such non BPS
dyons as a field theory configuration has been performed within
spherically symmetric ansatz and the resulting brane configurations
were found to agree with that of the type IIB string
theory\cite{sutcliffe}.  )
 
The detail of the 1/4 BPS field configurations has been explored in
Ref.~\cite{yi}. The BPS equations satisfied by these field
configurations consist of two pieces: one is the old magnetic BPS
equation for some 1/2 BPS monopole configuration while the other is a
covariant Laplace equation for the additional Higgs field in the 1/2
BPS monopole background.  Because of this two-tier structure of the
BPS equations, the parameter space of a 1/4 BPS dyon is identical to
the moduli space of the first, magnetic BPS equations. The only
subtlety here is that some of the monopole parameters transmute to
classical electric charges, as we compare the two.
 
The main lesson we learn from this fact is that the nonrelativistic
dynamics that incorporates 1/4 BPS dyonic states can be formulated as
that of dynamics on the same old monopole moduli space but with new
interactions. This is true, at least when the monopole rest mass is
dominant over the electric part of the energy.  The kinetic term of
the modified effective Lagrangian is given by the moduli space metric
of the underlying 1/2 BPS monopoles, while the potential term comes
from the additional Higgs field and is found to be a half of the
squared norm of a Killing vector.  The effective Lagrangian have a BPS
bound of its own. Classically and quantum mechanically, the 1/4 BPS
dyons arise as specific bound states that saturate such a low energy
BPS bound. This new dynamics incorporates the 1/4 BPS dyons as well as
more traditional 1/2 BPS monopoles and dyons.

There have been a couple of derivations of this effective low energy 
Lagrangian of monopoles that produces 1/4 BPS dyonic configurations. 
The first such derivation relied on the relation between the BPS energy  
of 1/4 BPS configurations and the conserved electric charges~\cite{blly}.  
The trick here is to realize that the BPS energy can be estimated in two 
different ways. One from the field theory, and the other from realizing 
1/4 BPS states as a bound state of monopoles in the low energy sense. By 
comparing the former, exact, formula to the latter, one can in fact 
identify the form of the potential term as one half of the electric  
part of the BPS energy, expressed in terms of Higgs expectation values  
and monopole moduli space geometry. The resulting potential 
is exact within the nonrelativistic approximation. The potential leads
to the long range attraction or repulsion between dyons~\cite{fraser}.
 
Shortly thereafter, there appeared an alternate derivation  
by two of the authors~\cite{blee}. 
Here, the field theoretic Lagrangian is calculated for a given initial field  
data, which are made of 1/4 BPS configuration and its field {\it velocity}.  
The Lagrangian, after integrated over the space, turns out to give 
the sum of minus rest mass and the kinetic parts which consists of 
quadratic and also linear terms in velocity of moduli  
coordinates. This is somewhat similar to the consideration of the zero mode  
dynamics of a particle with nonzero momentum.   
After shifting the cyclic coordinates related to conserved electric charges,  
one gets the low energy Lagrangian including the potential energy. 
 
In both of the previous derivations, the 1/4 BPS dyons were used as the  
convenient stepping stone that leads to the above low energy Lagrangian. 
However, the dynamics is really that of monopoles, which naturally produces  
1/4 BPS dyons as bound states. The effective Lagrangian was successfully 
understood because the states therein were understood very well by 
other means. This funny state of affairs begs for a question whether 
there exists a more fundamental derivation of the dynamics based only  
on the properties of 1/2 BPS monopoles.  
 
As we will see, there is indeed such a derivation. In particular,
since the method does not rely on BPS properties of monopole bound states, 
it is applicable to situations where bound states are typically non BPS. 
Such outcomes are generic 
when more than two Higgs take independent vacuum expectation values. 
In fact, we are going to find the exact low energy Lagrangian when all 
six Higgs are turned on. On the other hand, note that  
the low energy dynamics is meaningful only when the kinetic and the potential  
energies are much smaller than the rest mass of the monopoles~\cite{blly}.  
Because of this, one combination of Higgs field must be chosen to be large,  
so that monopoles have rest masses much larger than the electric and the  
kinetic part of energy. This separates six Higgs into one with large  
expectations and five remaining ones with small and independent expectation  
values.   
 
In this new derivation, the 1/2 BPS monopole configurations are primary  
and of order one. As we take the expectation values of the additional  
five Higgs fields to be small, we take them to be of order $\eta<<1$  
quantities, where $\eta$ characterizes the ratio of the additional Higgs  
expectation values to 
the first Higgs which shows up in magnetic BPS equation.  
We then solve the field equation for the five additional
Higgs fields classically 
to the leading order in $\eta$, given a static monopole background. The 
problem again reduces to a second order Laplace equation for each 
of five Higgs. We put the result back into the field theory Lagrangian to 
obtain the potential term as a function of Higgs expectations. Of course,  
independently of this, we also consider slow motions of monopoles and derive  
the kinetic term as well. The resulting action is accurate to  
the order, $\eta^2$ and $v^2$ where $v$ is the typical monopole velocity.

The modified effective Lagrangian is again based on the monopole moduli space, 
since, to the leading order, the above two computations of kinetic and
potential terms, do not interfere with each  
other. In particular, the kinetic  
term is given by the same moduli space metric. The potential is now half  
the sum of  squared norms of five Killing vectors. These five 
Killing vector fields are picked out by the five additional and small  
Higgs expectation values. For generic but small vev's of these five Higgs,  
even the lowest energy configuration with generic electric charges  
is non BPS.  
 
The view we take here is similar to the method of obtaining the 
Coulomb potential between two massive point particles. In that case, we 
consider the limit the electric coupling $e$ is small and the velocity 
$v$ of charged particles is small. Then, the leading solution of the 
Maxwell equation is Coulombic. To the order $v^2+ e^2$, the Lagrangian 
is the standard kinetic energy with the Coulomb potential. The 
retarded or relativistic effects would be of order $v^2 e^2, v^4$ and, 
thus, negligible. 
 
Recall that the planar string web for the 1/4 BPS dyonic
configurations are made of the webs of fundamental strings and D
strings. The key cause for the web is the attractive force between
fundamental and D strings. For the 1/4 BPS supersymmetry, the
orientation of the each string vertex should be consistent. In dyonic
picture, the 1/4 BPS configurations correspond to  dyons in finite
separation with the delicate balance between Higgs force and
electromagnetic force. They appear naturally as the BPS configuration
for the low energy dynamics. Thus they are the lowest energy
configurations for a given set of the electric charge, which
characterizes the BPS energy. In the non BPS case, again the non
planar string webs are formed from the fundamental strings and D
strings. In the low energy Lagrangian, they would correspond to the
lowest energy configuration for the given set of electric charges,
which could not saturate the BPS energy bound.
 
The supersymmetric completion of the low energy dynamics should have 
eight real  supercharges. For the cases with one Higgs expectation 
and two Higgs expectations, the supersymmetric low energy dynamics are 
known \cite{blum,blly,dongsu}. The former is completely determined 
by the monopole moduli space metric, which happens to be hyperK\"ahler,  
while the latter also involves additional potential determined by a  
single linear combination of triholomorphic Killing vector field on the 
moduli space. When all six Higgs fields are involved, the low energy 
dynamics involve up to five linearly independent combinations of 
such Killing vectors. The final goal of this paper is to write down 
this supersymmetric low energy Lagrangian explicitly\footnote{ The
supersymmetric quantum mechanics can be obtained as the dimensional   
reduction of the six dimensional (0,8)  supersymmetric sigma model
\cite{freed} to a one dimensional quantum   mechanics, by the
Scherk-Schwarz mechanism\cite{scherk}.   
This suggests that the form of the supersymmetric Lagrangian is the most  
general nonlinear sigma model with eight real supercharges. }, 
which completes the  low energy interaction of monopoles in $N=4$
Yang-Mills  theory, to the extent that the nonrelativistic
approximation makes sense.

The plan of the paper is as follows. In Sec. 2, we review the 1/2 BPS 
monopoles and the 1/4 BPS equations. In Sec. 3, we derive the 
effective Lagrangian for the non BPS configurations. In Sec. 4, we 
explore this Lagrangian. In Sec. 5, we find the  supersymmetric 
completion of the Lagrangian. In Sec. 6, we conclude with some 
remarks.

\section{BPS Bound and Primary BPS equation} 
 
We begin with the  ${ N}=4$  supersymmetric Yang-Mills  
theory. We choose the compact semi-simple group $G$ of the rank $r$. 
We divide the six Higgs fields into $b$ and $a_I$ with $I=1,...,5$. 
The  bosonic part of the Lagrangian is given  by 
\begin{eqnarray} 
L&=&  \frac{1}{2}  \int d^3x\; {\rm tr} \left\{ {\bf E}^2 +(D_0 b)^2+ 
(D_0a_I)^2 \right\} \nonumber \\ 
 &-&  \frac{1}{2} \int d^3x\; {\rm tr} \left\{   {\bf B}^2 + 
  ({\bf D} b)^2 +  ({\bf  D} a_I)^2+ \bigl( 
-i[a_I,b]\bigr)^2   + \sum_{I<J} \bigl(-i[a_I,a_J])^2 \right\}, 
\label{lag} 
\end{eqnarray} 
where $D_0 = \partial_0 -i A_0$, ${\bf D} = \nabla - i {\bf A}$, 
and ${\bf E} = \partial_0 {\bf A} - {\bf D} A_0$. 
The four vector potential $(A_0, {\bf A}) = (A_0^a T^a, {\bf 
A}^a T^a)$ and the group generators $T^a$ are traceless hermitian 
matrices such that $\tr T^a T^b=\delta^{ab}$.  
 
As shown in Ref.~\cite{yi},  there is a BPS bound on the energy  
functional, which is saturated only when  configurations satisfy 
\begin{eqnarray} 
&& {\bf B}={\bf D}b\,, 
\label{bogo1} \\ 
&&{\bf E}= c_I {\bf D} a_I \,,\ \ 
\label{bogo2} \\ 
&& 
D_0 b-{i}[c_I a_I, b]=0\,, \label{bogo3} \\ 
&&  
D_0 c_I a_I=0\,,\ \  \label{bogo4} 
\end{eqnarray}   
together with the Gauss law, 
\begin{equation} 
\label{gauss} 
{\bf D}\cdot {\bf E} -i[b,D_0 b]-i[c_Ia_I,D_0c_I a_I]=0\,.  
\end{equation} 
where $c_I$ is a unit vector in five dimensions. In addition, the rest 
of the Higgs field should be trivial on this configuration, or 
\begin{eqnarray} 
&& D_0 (a_I -c_I c_J a_J)=0 \\ 
&& {\bf D}(a_I - c_I c_J a_J ) = 0 \\ 
&& [b, a_I -c_I c_J a_J] = 0 \,. 
\end{eqnarray} 
This condition implies the 1/4 BPS configuration should be planar.  
The BPS energy is then 
\begin{equation} 
Z = {\bf b} \cdot {\bf g} + {c}_I {\bf a}_I \cdot {\bf q}\,, 
\end{equation} 
where  ${\bf b}$ and ${\bf a}_I$ are vacuum expectation values of the 
Higgs fields, while ${\bf g}$ and ${\bf q}$ are magnetic and electric
charges respectively.

Equation (\ref{bogo1}) is the old BPS equation for 1/2 BPS 
monopoles and is called the primary BPS equation. The BPS bound is 
saturated if the additional equations are satisfied.  
For 1/4 BPS configurations, the additional equations are from the 
energy bound and the Gauss law, which are put into a  
 single equation,  
\begin{equation} 
{\bf D}^2 c_Ia_I -[b,[ b, c_I a_I]]=0 ,  
\label{secondary} 
\end{equation}  
which is called the secondary BPS equation.  In addition, the $a_I-
c_I c_J a_J$ which is orthogonal to $c_I$ vector should commute with
all other fields and constant in space time. One last step necessary 
to solve for the 1/4 BPS dyon is to put $A_0 = - c_I a_I$. 
 
In the type IIB string realization of $U(N)$ Yang-Mills theories, the
above BPS equations imply that the corresponding 1/4 BPS string web
lies on a plane. However, the D3 branes which are not connected to the
web do not need to lie on the plane. Even when $D3$ branes lie on a
single plane, one can find  planar string webs which is not BPS as the
orientations of the string junctions are not uniform. In this paper, we
consider the special class of non BPS configurations which correspond
the non planar web, which would have been 1/4 BPS configurations when
we put  D3 branes to a single plane by small deformations of their
positions. In addition, we consider the string web is almost linear. 
For this case, we do not need to solve
the full quadratic field equations. We consider an
approximation by considering the quantity
\beq 
\eta \sim \frac{ |{\bf a}_I|}{| {\bf b}|}  ,
\eeq 
to be much smaller than unity, throughout this paper. 
 
Within such approximation, we may solve the field equation in two steps. 
First one solve the first-order, magnetic BPS equation. Once this is done, 
the solution of this primary BPS equation describes the collection of 
1/2 BPS monopoles, and the Higgs field $b$ takes the form 
\beq 
 b \simeq {\bf b}\cdot {\bf H} - \frac{{\bf g}\cdot {\bf H}}{4\pi r},  
\label{aasympt} 
\eeq 
asymptotically, where ${\bf H}$ is the Cartan subalgebra.  We are interested  
in the case where the expectation value ${\bf b}$ breaks the gauge group $G$ 
maximally to Abelian subgroups $U(1)^r$.  Then, there exists a unique set 
of simple roots $\betabf_1,\betabf_2,...,\betabf_r$ such that 
$\betabf_A \cdot {\bf b} > 0$~\cite{ejw}. The magnetic and 
electric charges are given by 
\begin{equation} 
 {\bf g} = 4\pi \sum_{A=1}^{r} n_A \betabf_A, 
\end{equation} 
where integer $n_A \ge 0$. For each simple root $\betabf_A$, there exist  
a fundamental monopole of magnetic charge $4\pi\betabf_A/e$, which comes 
with four bosonic zero modes: The integer $n_A$ can be thought of as the  
number of the $\betabf_A$ fundamental monopoles.  The moduli space of such  
1/2 BPS configurations has the dimension of $4\sum_A n_A$.  
We will consider the case where all $n_A$ are positive so that the monopoles  
do not separate into mutually noninteracting subgroups.  
 
Let us denote the moduli space coordinates by $z^m$.  If we
 parameterize BPS monopole solutions by the moduli coordinate $z$'s,
 $A_\mu({\bf x}, z^m) = ({\bf A}({\bf x}, z^m),b({\bf x}, z^m))$ with
 $\mu=1,2,3,4$, the zero modes are in general of the form,
\begin{equation} 
\delta_m A_\mu = \frac{ \partial A_\mu }{\partial z^m }+ D_\mu 
\epsilon_m, 
\label{deltaA}
\end{equation} 
where $D_\mu \epsilon_m = \partial_\mu \epsilon_m - i [A_\mu,
\epsilon_m]$ with understanding $\partial_4=0$. The zero modes around
the 1/2 BPS configurations are determined by perturbed primary BPS
equation plus a gauge fixing condition,
\begin{eqnarray} 
&& {\bf D} \times \delta_m {\bf A} = \nabla \delta_m b - i [ \delta_m {\bf 
A}, b] ,\\ 
&& D_\mu \delta_m A_\mu = 0, 
\label{background} 
\end{eqnarray} 
which forces the actual zero modes to be a sum of two terms. Given
this definition of   
zero modes, one can define a natural metric on the moduli space spanned 
by the collective coordinate $z$'s \cite{manton,atiyah,blum},  
\begin{equation} 
g_{mn}(z) = \int d^3 x \; {\rm tr} \:\delta_m A_\mu \delta_m A_\mu. 
\end{equation} 
With such a metric, the Lagrangian (\ref{lag}) for the monopoles of 
the primary BPS equation can be expanded for small velocities as, 
\begin{equation} 
\bar {\cal L}  = - {\bf g}\cdot {\bf b} + {\cal L} +\cdots   ,
\end{equation} 
where the first term is the rest mass of the monopoles, 
\begin{equation} 
{\bf g}\cdot {\bf b} =\frac{1}{2} \int d^3x \; \tr \left\{ {\bf B}^2 
  + ({\bf D} b)^2 \right\}\,. 
\eeq 
Ignoring the other five Higgs fields, the low energy dynamics that  
actually dictates the motion of these 1/2 BPS configurations would be  
given by the  purely kinetic, nonrelativistic Lagrangian 
\begin{equation} 
{\cal L} = \frac{1}{2} \,g_{mn}(z)\dot{z}^m \dot{z}^n\,. 
\label{lag12} 
\end{equation}

As there are $r$ unbroken global $U(1)$ symmetries, the corresponding
electric charges should be conserved. In other words, ${\cal L}$
should have $r$ cyclic coordinates corresponding to these gauge
transformations. In particular, we can choose a basis such that a
cyclic coordinate is denoted by $\xi^A$ ($ A=1,...,r $) corresponds to
the center of mass phase of monopoles of $\betabf_A$ root.  In
geometrical terms, the cyclic coordinates $\xi^A$'s generate Killing
vectors, 
\begin{equation} 
K_A\equiv\frac{\partial}{\partial \xi^A}\,. 
\end{equation} 
Finally, let us divide the moduli coordinates ${z^m}$ to ${\xi^A}$ and the 
rest $y^i$, upon which the Lagrangian (\ref{lag12}) can be rewritten as 
\begin{equation} 
{\cal L} = \frac{1}{2} h_{ij}(y) \dot{y}^i \dot{y}^j  
+ \frac{1}{2} L_{AB}(y) (\dot{\xi}^A + w^A_i(y) 
\dot{y}^i) (\dot{\xi}^B + w^B_j(y) 
\dot{y}^j)\,, 
\end{equation} 
which defines the quantities $h$, $L$, and $w$'s. In particular, 
\begin{equation} 
L_{AB} = g_{mn} K_A^m K_B^n\,. 
\end{equation} 
Notice that all metric components are independent of $\xi^A$. 
 
\section{Additional Higgs and Monopole Dynamics} 
 
Let us now explore the low energy dynamics of monopoles when
additional Higgs fields, $a_I$, are turned on. When 
 expectation
values ${\bf a}_I$ are turned on, the monopole solutions of the primary 
BPS equation are not, in general, solutions to the full field equations. 
Monopoles exert static forces on other monopoles. For sufficiently small
${\bf a}_I$, these forces arise from the extra potential energy due to 
nontrivial $a_I$ fields; The combined effect of ${\bf a}_I$ and of the 
monopole background induce some nontrivial behavior to $a_I$, which 
``dresses'' the monopoles and contributes to the energy of the system.

To find this  potential, we imagine a static configuration 
of monopoles, which are held fixed by some external force. Let us try to 
dress it with a time-independent $a_I$ field with the smallest 
possible cost of energy. The energy functional for such $a_I$ fields is
\begin{equation}
\Delta E=\frac{1}{2}\int d^3 x \; \tr\left\{    ( {\bf D} a_I)^2  
 +(-i\, [a_I,b\,])^2  \right\}  , \label{energy} 
\end{equation}
to the leading order where we ignore terms of higher power in $\eta$, 
such as $[\,a_I,a_J]^2$. We can find the minimal ``dressing'' field $a_I$
by solving the second order equation,
\begin{equation}
D^2 a_I - [b,[b,a_I]] = 0 .
\label{laplace}
\end{equation}
Solving this for $a_I$ and inserting them back into the energy functional
above, we should find the minimal 
cost of energy for 
the static monopole
configuration.

The same type of the second order equation appeared in construction
of 1/4 BPS dyons, where the projected Higgs field $c_I a_I$ obey such
an equation. However, we must emphasize that we are performing
a very different task here. Specifically, in the construction of 1/4 BPS
dyons, BPS equations force $- c_Ia_I $ to be identified with the 
time-component gauge field, $A_0$, which determines electric charges. 
Here we are simply solving for the reaction of the scalar fields $a_I$ 
to the given monopole configuration.

Using Tong's trick\cite{tong}, we notice that ${\bf D} a_I$ and $-i[b,a_I]$ 
can be thought of as  global gauge zero modes, $D_\mu a_I$,
which satisfy the gauge fixing condition, $D_\mu D_\mu a_I=0$. Thus,
$D_\mu a_I$ can be regarded as a linear combination of gauge zero modes, 
and subsequently each ${a}_I$ picks out  a linear combination
of $U(1)$ Killing vector fields on the moduli space, which are
\begin{equation} 
K^m_A \frac{\partial}{\partial z_m} = \frac{\partial}{\partial 
\xi_A}  .
\end{equation} 
More precisely, each $K_A$ corresponds to a gauge zero mode,
\begin{equation}
K_A^m \delta_m A_\mu,
\end{equation}
and each $D_\mu a_I$ is a linear combination of them,
\begin{equation} 
D_\mu a_I = a_I^A K_A^m \delta_m A_\mu, 
\end{equation} 
when we expand the asymptotic value ${\bf a}_I = \sum_A a^A 
\lambdabf_A$, where $\lambdabf_A$'s are the fundamental weights such 
that $\lambdabf_A \cdot \betabf_B = \delta_{AB}$.

We then express the potential energy ${\cal V}$, obtained by minimizing
the functional $\Delta E$ in Eq.~(\ref{energy}) in the monopole background, 
in terms of the monopole moduli parameters as
\begin{equation} 
{\cal V} = \frac{1}{2}\int d^3 x \:{\rm tr}\:\left\{(a_I^A K_A^m\delta_m A_\mu)
(a_I^B K_B^n\delta_n A_\mu)\right\}
=\frac{1}{2}\, g_{mn}\,a_I^A K_A^m  a_I^B  K_B^n.
\end{equation} 
The value of this potential depends on the monopole
configuration we started with, which induces the static force
on monopoles. The low energy effective Lagrangian was purely
kinetic when $a_I$ were absent. In the presence of $a_I$'s and of their
expectation values ${\bf a}_I$, however, the Lagrangian picks up a
potential term,
\begin{equation}
{\cal L}= \frac{1}{2}\,g_{mn}\dot z^m\dot z^n -{\cal V} 
\end{equation}
which can be written more explicitly as,
\begin{eqnarray}
{\cal L} &=& \frac{1}{2}  g_{mn}(z) \dot{z}^m \dot{z}^n 
-\frac{1}{2} g_{mn}(z) a_I^A K_A^m  a_I^B  
K_B^n\nonumber\\
&=&\frac{1}{2} h_{ij}(y) \dot{y}^i \dot{y}^j  
+ \frac{1}{2} L_{AB}(y) (\dot{\xi}^A + w^A_i(y) 
\dot{y}^i) (\dot{\xi}^B + w^B_j(y) 
\dot{y}^j)-{1\over 2} L_{AB}(y) a^A_I a_I^B \,. 
\label{lag1} 
\end{eqnarray} 
where the index $I$ runs from 1 to 5, and labels the five 
potential terms.

The procedure we employed here should be a very familiar one. When we
talk about, say, Coulombic interaction between charged particles, we 
also fix the charge distribution by hand, and then estimate the potential
energy it costs. Of course, there is a possibility of more interaction terms
involving velocities of moduli as well as $a_I$ fields, but in the low
energy approximation here, the only relevant terms of such kind would be 
of order $v\eta$. However, it is clear that neither backreaction of $a_I$
to the magnetic background nor the time-dependence of $a_I$'s can 
produce
such a term. Thus, to the leading quadratic order in $v$ and $\eta$, the 
above Lagrangian captures all bosonic interactions among monopoles
in the presence of ${\bf a}_I$'s.\footnote{While the low energy dynamics
turns out to be quite simple, there is a subtlety in reconstructing the 
actual field configuration 
for a given low energy motion on the moduli space. For the magnetic part of the 
configuration, $A_\mu$, the trajectory on moduli space can be 
represented reliably  by allowing time-dependence of the moduli parameters.
Namely, the time-dependent field configuration would be 
$A_\mu=\tilde A_\mu({\bf x};z_m(t))$ where $\tilde A_\mu({\bf x};z)$
is the solution of the primary BPS equation. For the additional Higgs fields, 
however, the naive ansatz $a_I=\tilde a_I({\bf x}; z_m(t))$ does not work,
where $\tilde a_I({\bf x}; z_m)$
solves the static second order equation (\ref{laplace}) in the background
of $\tilde A_\mu({\bf x};z)$. Such an ansatz
would involve fluctuations of nonnormalizable modes, as $\tilde a_I$ has
a $z_m$-dependent $1/r$ tail. Rather, the time-dependence of $a_I$ field
has much nicer large $r$ behavior, and this can be seen easily by
solving the full field equation for $a_I$ order by order in $v$.}

\section{1/4 BPS and Non BPS Configurations} 
 
The total energy of the field configuration within this nonrelativistic 
approximation is then 
\begin{equation} 
E = {\bf b}\cdot {\bf g} + {\cal E} \,, 
\end{equation} 
where the nonrelativistic energy is derived from $\cal L$, and can be 
written as 
\begin{equation} 
{\cal E}=  \frac{1}{2}\, g_{mn}(z) \left( 
\dot{z}^m \dot{z}^n +a_I^A K^m_A a_I^B K_B^n\right)\,. 
\end{equation} 
The energy ${\cal E} $ has a BPS bound of its own. With an arbitrary  
five dimensional unit vector $c_I$, we can rewrite the energy as 
\begin{equation} 
{\cal E} = \frac{1}{2} g_{mn} ( \dot{z}^m - c_I a_I^A K_A^m) 
(\dot{z}^n - c_J a_J^B K_B^n) + \frac{1}{2} 
g_{mn}{a}_{\bot I}^A K_A^m {a}_{\bot J}^B K_B^n + c_I g_{mn} 
\dot{z}^m a_I^A K_A^n 
\end{equation} 
where ${a}_{\bot I}^A = a_I^A - c_I c_J a_J^A$ is the part of $a_I^A$ 
orthogonal to $c_I$. 
Since there is $r$ $U(1)$ symmetries with Killing vectors $K_A^m$, 
there are $r$ conserved charges  
\begin{equation} 
q_A = K_A^m \frac{\partial {\cal L}}{\partial \dot{z}^m} =  
g_{mn} K_A^m \dot{z}^n \,. 
\end{equation} 
As the metric $g_{mn}$ are positive definite, there is a bound on the 
energy, 
\begin{equation} 
E \ge | c_I a_I^A q_A |\,. 
\end{equation} 
This bound is saturated when  
\begin{eqnarray} 
&&  \dot{z}^m - c_I a_I^A K_A^m = 0 \\ 
&&   a_{\bot I}^A=a_I^A - c_I c_J a_J^A = 0 \,. 
\end{eqnarray} 
The second  equation is satisfied if, for instance, only 
one additional Higgs fields are relevant, while the first equation implies  
that the conserved charges are 
\begin{equation} 
q_A = g_{mn}   K_A^m c_I a_I^B     K_B^n\,. 
\label{bpscharge} 
\end{equation} 
Quantum counterpart of such BPS configurations have been explored  
in Ref.~\cite{dongsu}. In field theory terms, these BPS states of 
low energy dynamics preserve 1/4 of field theory supersymmetries. 
 
These 1/4 BPS configurations describe static dyons spreading out in
space such that the electromagnetic force and the Higgs force are in
delicate balance. They are the BPS configurations of the low energy
effective action when, in effect, only one linear combination of the
Killing vector fields, $c_I a_I^A K_A$, is relevant.  For more general
cases, when $a_{\perp I}$ cannot be taken to be zero, the BPS bound
are not saturated. Nevertheless, there must exist the lowest energy
state with any given charge, which would correspond to stable non BPS
states. (These non BPS configurations correspond to the string web
which is not planar.) Such a stable dyonic configuration can be found
classically considering the nonrelativistic energy functional.
 
The energy functional for a given set of electric charges 
\begin{equation} 
q_A = \frac{\partial {\cal L}}{\partial \dot{\xi}^A} 
\end{equation} 
is 
\begin{equation} 
{\cal E}= \frac{1}{2} h_{ij}\dot{y}^i\dot{y}^j + U_{\rm eff}(y) 
\end{equation} 
where the effective potential is 
\begin{equation} 
U_{\rm eff} = \frac{1}{2} L^{AB}(y) q_A q_B + \frac{1}{2}  
L_{AB} a_I^A a_I^B 
\end{equation} 
with the inverse of $L_{AB}$ is denoted by $L^{AB}$. 
The minimum of the energy is achieved by the configurations which are 
static in $y^i$ and satisfy 
\begin{equation} 
\frac{\partial}{\partial y^i} U_{\rm eff}(y)=0 
\label{crit} 
\end{equation} 
In general the family of stable solutions {$y^i$} for a given $q_A$,
if they exist, will form a submanifold of the moduli space. However,
it is not clear whether there will be always $q_A$ satisfying
Eq. (\ref{crit}) for some $y^i$. In fact, it is known that for some
case with too large values of $q_A$ there is no solution to such
equations\cite{yi}.
 
The general analysis of Eq.~(\ref{crit}) will be complicated. One case 
where it can be solved explicitly is when the magnetic background contains 
only one fundamental monopole of each kind; That is, suppose that, for  
each simple root $\betabf_A, A = 1,...,r$, we have one fundamental monopole  
at ${\bf x}_A$ and with the $U(1)$ phase $\xi_A$. Denote the  
relative position vectors between adjacent (in the Lie algebra sense)  
monopoles by ${\bf r}_A = {\bf x}_{A+1}-{\bf x}_A$ for $A=1,..., r-1$  
and also define the corresponding relative phases by $\zeta_A$.  
For the phases, the redefinition is such that the charges $\tilde q_A$,  
associated with $\xi_A$, is related to $q_A$'s by $\tilde q_A= 
q_{A+1}-q_A$. The metric is then decomposed into two decoupled 
pieces\cite{weinbergyi}; 
\begin{eqnarray} 
ds^2 &=&\frac{1}{(\sum m_A)}\left(  
d(\sum_{A=1}^r m_A {\bf x}_A)^2 +\frac{16\pi^2}{e^4} 
d(\sum_{A=1}^r\xi_A)^2\right)\nonumber \\ 
&& \nonumber \\ 
&+& 
\sum_{A=1}^{r-1}\sum_{B=1}^{r-1} \left( C^{AB} d{\bf r}_A \cdot d{\bf r}_B +  
C_{AB}(d\zeta^A + {\bf w}({\bf r}_A)\cdot d{\bf r}_A)(d\zeta^B + {\bf 
w}({\bf r}_B) \cdot 
d{\bf r}_B)  \right)
\end{eqnarray} 
where $m_A$ are the masses of the $r$  fundamental monopoles. The 
$(r-1)\times (r-1)$ matrices $C^{AB}$ and $C_{AB}$ are inverses of each 
other, 
\begin{equation} 
\sum_{B=1}^{r-1} C^{AB}C_{BC}=\delta^A_C 
\end{equation} 
and are explicitly known 
\begin{equation} 
C^{AB} = \mu^{AB} + \delta^{AB}\frac{\lambda_A}{|{\bf r}_A|} 
\end{equation} 
with the reduced mass matrix $\mu_{AB}$, $A,B=1,\dots,r-1$, and some  
coupling constants $\lambda_A$. The vector potential ${\bf w}({\bf r})$  
is the Dirac potential; 
\begin{equation} 
\nabla \frac{1}{r} = \nabla\times {\bf w}({\bf r}).  
\end{equation} 
In the new coordinate, the potential also decomposes into two parts, one 
of which is independent of moduli coordinates, 
\begin{eqnarray} 
U_{\rm eff}(r^A) &=& \frac{1}{2}\sum_{A,B=1}^r L^{AB}(y) q_A q_B 
+ \frac{1}{2}  
\sum_{A,B=1}^r L_{AB} a_I^A a_I^B \nonumber \\  
&=&{\rm constant} + {1\over 2} \sum_{A,B=1}^{r-1}C^{AB} \tilde q_A  
\tilde q_B  +  {1\over 2} \sum_{A,B=1}^{r-1} C_{AB} \tilde a_I^A \tilde a_I^B 
\label{distinctpotential} 
\end{eqnarray} 
The vacuum expectation values in the new basis, $\tilde a_I^A$,  
$A=1,\dots,r-1$, are found from $a_I^A$, $A=1,\dots, r$, using the
relationship,
\begin{equation}
\sum_{A=1}^r a_I^Aq_A =\sum_{A=1}^{r-1}\tilde a^A_I \tilde q_A + \tilde a_I^0
\frac{\sum_{A=1}^r m_A q_A}{\sum_{A=1}^r m_A}
\end{equation}
with $\tilde a_I^0$ to be determined from this as well.

The minimum of the potential is found by looking for the critical point, 
\begin{equation} 
0=\frac{\partial}{\partial {\bf r}_C} U_{\rm eff} = \sum_{A,B=1}^{r-1} 
\frac{\partial C^{AB}}{2\partial {\bf r}_C} \left( \tilde q_A \tilde q_B -  
\sum_{I=1}^5\sum_{A',B'=1}^{r-1}C_{A A'} C_{B B'} 
\tilde a_I^{A'} \tilde a_I^{B'} \right) 
\end{equation} 
As $\partial C^{AB}/ \partial {\bf r}_C = -\delta_{AB}\delta_{BC}  
{\bf r}_C/(r_C)^3$, the condition reduces to 
\begin{equation} 
0=\frac{\partial}{\partial {\bf r}_C} U_{\rm eff} = - 
\frac{\lambda_C{\bf r}_C}{r_C^3 } \left( (\tilde q_C )^2  
- \sum_{I=1}^5\sum_{A,B=1}^{r-1} 
C_{C A} C_{CB} \tilde a_I^{A} \tilde a_I^{B} \right) 
\end{equation} 
and we find  that the critical points are such that the charges are given
as functions of $\vec r_A$ as follows,
\begin{equation} 
|\tilde q_C| = \sqrt{\sum_{I=1}^5 \sum_{A,B=1}^{r-1}C_{CA}(\vec r) 
C_{CB}(\vec r)\, \tilde a_I^{A} \tilde a_I^{B}} 
\label{nonbpscharge} 
\end{equation} 
Once this is satisfied, $\dot{\bf r}_A=0$ solves the equations of motion, 
so the solution describes static configurations of many distinct monopoles, 
each dressed by the electric charges. They corresponds to stable   
non BPS dyons in the field theoretic description.    
 
By inserting (\ref{nonbpscharge}) to the effective potential  
(\ref{distinctpotential}), the energy of the configuration 
is determined as a function of the monopole positions.    
 The latter $U(1)$ charge is not determined 
by moduli parameters. It should be remarked that these states become
1/4 BPS, when only one Higgs  vacuum expectation value out of five is 
nonvanishing or all the directions of them are parallel with each other.

\section{Supersymmetric Extension}  
  
So far we have  concentrated on the bosonic part of  
the low energy effective Lagrangian in the moduli space   
approximation. The supersymmetric extension of the bosonic  
action can be achieved in two different routes. A direct   
approach is to follow the same strategy of the bosonic part.  
Namely, identify first the moduli fluctuations and their   
coordinates of the fermionic part, and integrate    
out all the other fluctuation except the monopole moduli variables   
using the original field theoretic Lagrangian.  

While such a derivation would be more desirable, the symmetry of
the system seems to offer us a shortcut and allow us to fix the SUSY
completion of the effective Lagrangian uniquely. We shall follow  
the latter approach here. 

Since the background configurations of monopoles preserves half of the
16 supersymmetries of the original ${N}$=4 SYM theory, the quantum
mechanics should have four complex, or eight real supercharges,
Furthermore, the low energy effective theory should be consistent with
the $SO(6)$ R-symmetry of the four dimensional ${N}$=4 SYM theory.
Out of six Higgs fields, we picked out one, $b$, associated with
construction of monopoles, so only $SO(5)$ subgroup of $SO(6)$ may
show up. For instance, when we consider the extreme case of ${\bf
a}_I=0$, the SUSY quantum mechanics must have full $SO(5)$ R-symmetry.

The additional Higgs expectations ${\bf a}_I$ break the remaining $SO(5)$ 
rotational symmetry of the field theory, as well, spontaneously.
On the other hand, in the low energy dynamics of monopoles, ${\bf a}_I$ 
are small parameters, so the breaking of $SO(5)$ is explicit and soft.
Thus, $SO(5)$ is not a symmetry of the low energy dynamics. Nevertheless,
because all ${\bf a}_I$ are on an equal footing (unlike $\bf b$), the low 
energy dynamics must remain invariant when we rotate the ${\bf a}_I$ 
in addition to rotating dynamical degrees of freedom. Thus,
although this $SO(5)$ is not a symmetry of the low 
energy theory in the conventional sense, this provides us with an 
interesting consistency checkpoint. Later we will find and write down 
this $SO(5)$ transformation explicitly.
 
Existence of the four complex supercharges is already quite restrictive.
The supersymmetry requires the geometry to be hyperK\"ahler, to begin with,
equipped with three complex structures that satisfy
\beqn  
&&{\cal I}^{(s)}{\cal I}^{(t)}  =  
- \delta^{st} +\epsilon^{stu} {\cal I}^{(u)}, \\  
&& D_\bmu {\cal I}^{(s)\bnu}\,_\bp  =0\, .  
\label{complexstructure}  
\eeqn  
In the absence of ${\bf a}_I$'s, the dynamics would be a sigma model
onto the hyperK\"ahler moduli space of monopoles. The bosonic potential
introduced by ${\bf a}_I$'s can be rewritten in terms of five triholomorphic
Killing vector fields, $G_I\equiv {\bf a}_I\cdot{\bf K}$, as
\begin{equation}
\frac{1}{2}\sum_{I=1}^5 G_I^m G_I^n g_{mn} .
\end{equation}

Alvarez-Gaume and Freedman \cite{gaume} discussed how such
Killing vector fields can be incorporated into supersymmetric Lagrangian 
while maintaining
four complex supercharges 
in the two-dimensional context.
In this two-dimensional setting, they showed that up to four triholomorphic
Killing vectors can be accommodated. This result presumably has something
to do with the fact that the supersymmetric Lagrangian can also be obtained
via the Scherk-Schwarz dimensional reduction\cite{scherk} from the six 
dimensional (0,8) nonlinear sigma model action presented in Ref.~\cite{freed}.

Since we are considering quantum mechanics instead of two-dimensional 
field theory, this suggests to us that one should be able to incorporate 
up to five such Killing 
vectors to the effective Lagrangian. Thus, generalizing their result
to quantum mechanics, we obtain the following unique supersymmetric
completion of the low energy dynamics,
\beqn  
{\cal L}={1\over 2} \biggl( g_{\bmu\bnu} \dot{z}^\bmu \dot{ z}^\bnu +  
ig_{\bmu\bnu} \bar\psi^\bmu \gamma^0 D_t \psi^\bnu + {1\over 6}  
R_{\bmu\bnu\bp\bq}\bar\psi^\bmu \psi^\bp \bar\psi^\bnu \psi^\bq  
- g_{\bmu\bnu} G_{\bI}^\bmu   
G_{\bI}^\bnu  
-i D_\bmu G_{\bI \bnu} \,\, \bar\psi^\bmu  
(\Omega^\bI \psi)^\bnu  
\biggr)  
\label{susyaction}  
\eeqn  
where $\psi^\bmu$ is a two component Majonara spinor, $\gamma^0=\sigma_2, 
\gamma^1=i\sigma_1, \gamma^2=-i\sigma_3$, $\bar\psi=\psi^T\gamma^0$. 
The operator $\Omega_\bI$'s are defined respectively by 
$\Omega_4=\delta^\bmu_\bnu \,\gamma^1_{\alpha\beta}$, 
$\Omega_5=\delta^\bmu_\bnu \,\gamma^2_{\alpha\beta}$ and 
$\Omega_s=i{\cal I}^{(s)\bmu}\!\,_\bnu \delta_{\alpha\beta}$ for  
$s=1,2,3$. 
 
The supersymmetry algebra by itself requires some properties of $G_I$'s,
in addition to hyperK\"ahler properties of $g_{mn}$.
$G_{\bI}$ must satisfy  
\beqn 
D_\bmu G_{\bI \bnu} +D_\bnu G_{\bI \bmu}  =0\,, 
\eeqn 
or  
equivalently ${\cal L}_{G_{\bI}} g=0$, that is, $G_I$ must be 
a Killing vector. In addition, the rotated version $({\cal I}^{(s)} 
G_{\bI})_m$'s must also satisfy  
\beqn 
D_\bmu ({\cal I}^{(s)}G_{\bI})_\bnu- 
D_\bnu ({\cal I}^{(s)}G_{\bI})_\bmu=0 
\eeqn 
Taken together with the Killing properties of $G_I$ and the closedness
of the K\"ahler forms, this also implies that $G_I$ are triholomorphic,
\begin{equation}
{\cal L}_{G_{\bI}} {\cal I}^{(s)}=0
\end{equation}
Of course, for the specific case of monopole dynamics, these two conditions
are satisfied because each $K_A$ is a triholomorphic Killing vector field
on the moduli space. One last requirement on $G_I$'s from SUSY algebra is,
\begin{equation} 
G_{\bI}^{\bmu}\,{\cal I}^{(s)}_{\bmu\bnu}\, G_{\bJ}^{\bnu}=0 .
\end{equation}
for $s=1,2,3$.
This condition is met for triholomorphic $G_I$'s, provided that the 
commutators vanish, $[G_I,G_J]=0$. Since $K_A$'s all commute among themselves,
this last condition is also satisfied in the above low energy dynamics of
monopoles. 
 
When quantized,  
the spinors $\psi^E = e_\bmu^E \psi^\bmu$ with vielbein $e_\bmu^E$,  
commute with all the bosonic dynamical variables, especially with  
$p$'s that are canonical momenta of the coordinates $z$'s. 
The remaining fundamental  commutation relations are  
\beqn  
&&[z^\bmu, p_\bnu ] = i\delta^\bmu_\bnu, \nonumber\\  
&&\{\psi^E_\alpha, \psi^F_\beta\} = \delta^{EF}\delta_{\alpha\beta}\,.  
\label{commutators}  
\eeqn  
 (Consequently, the bosonic momenta $p$'s do not  
commute with $\psi^\bmu$.) It is straightforward to show that 
the Lagrangian (\ref{susyaction}) is invariant under the N=4  
supersymmetry transformations,  
\beqn  
&&  
\delta_{(0)} z^\bmu= \bar\epsilon \psi^\bmu, \nonumber\\  
&& \delta_{(0)}\psi^\bmu  = -i\dot{z}^\bmu\gamma^0  
\epsilon  - \Gamma^\bmu_{\bnu\bp}  
\bar\epsilon \psi^\bnu \psi^\bp   
-i (G^{\bI}\Omega^\bI)^\bmu\epsilon, \\  
&&\delta_{(s)}z^\bmu  =  
{\bar\epsilon}_{(s)} ({\cal I}^{(s)}\psi)^{\bmu} ,\nonumber\\  
&& \delta_{(s)}\psi^{\bmu}\!\!  =  
\! i({\cal I}^{(s)}\dot{z})^\bmu\gamma^0\,  
\epsilon_{(s)} \! +\!{\cal I}^{{(s)} m}\!\,_l \,\Gamma^l_{\bnu\bp}  
\bar\epsilon_{(s)} ({\cal I}^{(s)}\psi)^\bnu  
({\cal I}^{(s)}\psi)^{\bp}   
-i(G^{\bI}{\cal I}^{(s)}\Omega^\bI)^\bmu\epsilon_{(s)}\,,  
\label{transformation}  
\eeqn  
where $\epsilon$ and $\epsilon_{(s)}$ are  spinor parameters 
and no summation convention is used for the index $s=1,2,3$. 
 
For the supercharges, let us  first  
define  supercovariant momenta by  
\begin{eqnarray}  
&& \pi_\bmu \equiv p_\bmu -{i\over 2}\omega_{EF\,\bmu}  
\bar\psi^E \gamma^0 \psi^F,  
\label{cov}  
\end{eqnarray}  
where $\omega_{EF\,\bmu}$ is the spin connection.  
The corresponding N=4 SUSY generators in real spinors  are then  
\begin{eqnarray}  
&&Q_\alpha = \psi^\bmu_\alpha \pi_\bmu  
 - 
(\gamma^0\Omega^\bI \psi)^\bmu G^{\bI}_\bmu, 
\label{generator0} 
\\  
&&Q^{(s)}_\alpha = ({\cal I}^{(s)}\psi)^{\bmu}_\alpha \pi_\bmu  
- 
(\gamma^0 {\cal I}^{(s)}\Omega^\bI \psi)^\bmu G^{\bI}_\bmu\,.  
\label{generator}  
\end{eqnarray}  
These charges satisfy the N=4 superalgebra:  
\beqn  
&&\{Q_\alpha,Q_\beta  \}  =\{Q^{(s)}_\alpha,Q^{(s)}_\beta\}=2  
 \delta_{\alpha\beta} \; {\cal H}  
-2(\gamma^0\gamma^1)_{\alpha\beta} \; {\cal Z}_{4} 
-2(\gamma^0\gamma^2)_{\alpha\beta} \; {\cal Z}_{5}, \\  
&& \{Q_\alpha,\ Q^{(s)\,}_\beta  \} =  
2\gamma^0_{\alpha\beta} \; {\cal Z}_{s},\ \ \  
\{Q^{(1)}_\alpha,Q^{(2)}_\beta  \} = 
2\gamma^0_{\alpha\beta} \; {\cal Z}_{3},\\  
&& \{Q^{(2)}_\alpha,Q^{(3)}_\beta  \} = 
2\gamma^0_{\alpha\beta} \; {\cal Z}_{1}, 
\ \ \  \{Q^{(3)}_\alpha,Q^{(1)}_\beta  \} = 
2\gamma^0_{\alpha\beta} \; {\cal Z}_{2}\,, 
\label{algebra}  
\eeqn  
where the Hamiltonian $\cal H$ and the central charges ${\cal Z}_I$  
read  
\beqn  
&&{\cal H}=  
{1\over 2} \biggl( {1\over \sqrt{g}}\pi_\bmu \sqrt{g }g^{\bmu\bnu}\pi_\bnu  
+ g^{\bmu\bnu}G^{\bI}_\bmu G^{\bI}_\bnu  
-{1\over 4}R_{\bmu\bnu\bp\bq}\bar\psi^\bmu \, 
\gamma^0 \psi^\bnu \bar\psi^\bp  
\gamma^0 \psi^\bq  
+i D_\mu G^{\bI}_\nu \, \bar\psi^\mu  
\Omega^\bI \psi^\bnu\biggr),\\  
&& {\cal Z}_{I}= G_{\bI}^{\bmu} \pi_\bmu - 
{i\over 2}  D_\bmu G^{\bI}_\bnu \,\bar\psi^\bmu  
\gamma^0\psi^\bnu.  
\label{hamiltonian}  
\eeqn  
It is straightforward to see that the SO(5) rotation is realized 
 by the 
 transformation, 
\beqn  
&&\psi \rightarrow  
e^{{1\over 2}\theta_{KL}J_{KL}}\,\,\psi\,, 
\label{phase1} 
\eeqn  
where $\theta_{KL}$ is antisymmetric in its indices and  
the corresponding  generators, $J_{KL}\,(\,=-J_{LK})$, denote 
\beqn  
&& J_{ab}=\epsilon_{abc}\,{\cal I}^{(c)}\,, \ \ \  
J_{45}=i\sigma_2\,, 
\ \ \ J_{4a}=\sigma_1 \,{\cal I}^{(a)}\,,\ \ \   
J_{5a}=-\sigma_3 \,{\cal I}^{(a)}  
\label{generator1} 
\eeqn  
with $a,b,c=1,2,3$. For example, the transformation reads 
explicitly 
\beqn  
\psi^\bmu_\alpha\ \ \rightarrow \ \ \cos{\theta} \,\psi^\bmu_\alpha + 
\sin{\theta} \, (\sigma_1 {\cal I}^{(1)}\psi)^\bmu_\alpha\,, 
\label{phase2}   
\eeqn  
when only $\theta_{41}=-\theta_{14}=\theta$ is nonvanishing. 
 
Performing such  SO(5) rotations,  we obtain a  theory 
with the vacuum expectation values ${{\bf a'}}_{\bI}={\cal R}_{\bI\bJ} 
{\bf a}_{\bJ}$ where ${\cal R}_{\bI\bJ}$ is the corresponding  
SO(5) rotation  
matrix satisfying ${\cal R}^T {\cal R}=I$. More specifically, 
the induced transformation of $G_I$ that is linear in  
the vacuum expectation  
value ${\bf a}^I$, is  
\beqn  
G_{\bI}\rightarrow \Bigl(e^{\theta_{KL} 
{\cal J}_{KL}}\Bigr)_{\bI\bJ} G_{\bJ} 
\label{phase11}  
\eeqn 
where $({\cal J}_{KL})_{IJ}={1\over 2}(\delta_{KI}\delta_{LJ}- 
\delta_{KJ}\delta_{LI})$.  
When all ${{a}}^{\bI}$ 
are parallel with each other, 
one may  
make only one Higgs expectation value nonvanishing by an appropriate  
SO(5) rotations; the result  
 corresponds to 1/4 BPS  
effective Lagrangian in Ref.~\cite{yi}.

The ten generators of SO(5) in (\ref{generator1}) exhaust all the possible 
covariantly constant, antisymmetric structures present in the N=4 
supersymmetric sigma model without potential, so the  
realization of the R-symmetry is rather unique. 
 
The complex form of the supercharges 
are often useful. For this, we introduce  
$\varphi^\bmu\equiv{1\over \sqrt{2}} (\psi_1^\bmu-i\psi_2^\bmu)$ 
and define $Q\equiv {1\over \sqrt{2}}(Q_1-iQ_2)$. The supercharges 
in (\ref{generator0}) can be rewritten as 
\begin{eqnarray}  
&&Q = \varphi^\bmu \pi_\bmu  
 -\varphi^{*\bmu} (G^4_\bmu-iG^5_\bmu)-i\sum_{s=1}^3 
G^s_\bmu ({\cal I}^{(s)}\varphi)^\bmu 
\\  
&& 
Q^\dagger = \varphi^{*\bmu} \pi_\bmu  
 -\varphi^{\bmu} (G^4_\bmu+iG^5_\bmu)+ 
i\sum_{s=1}^3 G^s_\bmu ({\cal I}^{(s)}\varphi^*)^\bmu 
\,. 
\label{generators}  
\end{eqnarray} 
The supercharges $Q^{(s)}$ and ${Q^{(s)}}^\dagger$ are 
found similarly from
(\ref{generator}).  
Finally, $\{Q,Q^\dagger\}= 
\{Q^{(s)},{Q^{(s)}}^\dagger\}={\cal H}$, so 
the Hamiltonian is positive definite. All the central charges  
appear in other parts of the algebra.  
 
In Ref. \cite{dongsu}, the quantum 1/4 BPS wavefunctions have  
been constructed explicitly  and the structure of the supermultiplet 
are identified. This construction has heavily relied upon 
the BPS nature of the states which
first order equations. This kind of simplification does not 
occur in the case of the stable non BPS states, so we will  
leave the analysis of their wavefunctions for future works.

\section{Conclusion} 
 
We have found the complete supersymmetric Lagrangian for the low energy 
dynamics of 1/2 monopoles when all six adjoint Higgs fields get expectation
values. We consider the nonrelativistic dynamics of monopoles, which 
constrains the five additional Higgs to be small compared the first that
gives mass to the monopoles. The bosonic part of the effective dynamics is 
found by a perturbative expansion of fields around purely magnetic monopole 
configurations, which agrees with previously found exact Lagrangian 
when only two Higgs fields, one large and another small, are involved. 
The supersymmetric extension 
is constructed and argued to be unique, given the four complex supercharges,
and an would-be $SO(5)$ R-symmetry that is softly broken by five Killing 
potential terms.

The dyonic state from the low energy dynamics would correspond to
a web of strings ending on D3 branes, when realized as type IIB
string theory configurations. This is possible for all classical 
gauge groups of the Yang-Mills theory. When the transverse positions of 
the D3 branes are planar, only one of the five Killing vectors
become relevant, and the state saturate a BPS bound which is linearly
characterized by the values of electric charges. These are 1/4 BPS dyons
in the field theory sense.

For a non planar distribution of D3 branes, on the other hand, at least two 
of the five Killing vectors are relevant. The resulting non planar web 
does not saturate a BPS bound, but exist as a stable dyonic state. The
state will consist of several distinct monopoles, each dressed with 
some electric charges that are mostly determined by the inter-monopole 
separations. For a simple case, we gave a set of algebraic equations
that can be used to determine the charge-position relationship.
  
We have not fully explored the low energy effective Lagrangian even
classically. It is expected that there exists a clear correspondence
between the energy of non planar string web and the minimum energy of
the stable but non BPS states. The energy of the string web in an
appropriate limit can be determined as sum of each length multiplied
by tension. The detailed comparison will be of interest.  It would be
interesting to find out the field theoretic configuration for these
dyonic non BPS configurations.  The quantum mechanics of the
supersymmetric Lagrangian is more involved than that of
Ref.~\cite{dongsu}, which considered only one Killing
potential. Nevertheless it is of some interest to find the ground
state for the given electric charges.

\centerline{\bf Acknowledgments} 
 
D.B. is supported in part by Ministry of Education Grant 
98-015-D00061.  K.L. is  supported in part by the SRC program of the 
SNU-CTP and the Basic Science and Research Program under BRSI-98-2418. 
D.B. and K.L. are also supported in part by KOSEF 1998 
Interdisciplinary Research Grant 98-07-02-07-01-5.

\end{document}